\documentclass[12pt]{iopart}

\usepackage{graphicx}

\begin{document}
\title{Sudden switching in two-state systems}
\author{Kh Kh Shakov$^1$, J H McGuire$^1$, L Kaplan$^1$, D Uskov$^2$\footnote{Permanent address: Lebedev Physical Institute, Moscow, Russia} and A Chalastaras$^1$}
\address{$^1$ Physics Dept., Tulane University, New Orleans, LA 70118, USA}
\address{$^2$ Physics Dept., Louisiana State University, Baton Rouge, LA 70803, USA}
\ead{kshakov@tulane.edu}
\begin{abstract}
Analytic solutions are developed for two-state systems (e.g. qubits) strongly perturbed by a series of rapidly changing pulses, called `kicks'.
The evolution matrix may be expressed as a time ordered product of evolution matrices for single kicks. 
Single, double, and triple kicks are explicitly considered, and the onset of observability of time ordering is examined. 
The effects of different order of kicks on the dynamics of the system are studied and compared with effects of time ordering in general.
To determine the range of validity of this approach, the effect of using pulses of finite widths for $2s-2p$ transitions in atomic hydrogen is examined numerically.   
\end{abstract}
\submitto{\jpb}

\maketitle

\section{Introduction}
For many years a wide variety of physical systems have been described, often approximately, 
in terms of coupled two-state systems~\cite{ae,shore,josephson}.  
In more recent years application has been found in quantum information and 
quantum computing~\cite{qcomp00}, where such two-state systems have been used
to describe a quantum mechanical version of the classical computer bit.
This quantum bit, or `qubit', is described as a linear superposition
of two states (say `off' and `on'), so that before a measurement the qubit is 
in some sense simultaneously both off and on, unlike a classical bit
which is always either off or on.

While two-state quantum systems are widely used, their utility suffers because there are only a limited number of known analytic solutions.
For most two-state systems numerical calculations are required.
Although standard numerical methods are both fast and reliable for these simple systems, finding how the corresponding physical systems work is largely a numerical fishing trip in cloudy waters.  
Analytic solutions, where they exist, are more transparent.
One such solution that has received widespread applications in quantum optics~\cite{ae,shore,mw} is obtained using the rotating wave approximation (RWA). 
In this approach, periodic transfer of the population within the system is achieved by applying an external field (e.g. laser) that is tuned to a narrow band of frequencies to match a particular transition between the system's levels. 
The result is well known Rabi oscillations.
In this paper we wish to call attention to another analytic solution for two-state quantum equations, namely the limit of sudden pulses.  
Such a fast pulse is called a `kick'.  
Unlike a periodic field with well defined carrier frequency used in the RWA technique, a kick is localized in time and consists of a broad range of frequency components.
Due to their 'non-periodic' structure, kicks are well suited for applications that require occasional modifications of the system's state (one kick - one transition). 
There are advantages in using kicks in systems where two states of interest lie close to one another in energy (nearly degenerate systems). 
And, of course, kicks would be a natural choice in cases where a pulsed source has to be used for one reason or another.
Fast pulses are an essential ingredient in the kicked rotor or standard map, a paradigm of the transition to chaos in one-dimensional time-dependent dynamics~\cite{kickedrotor}. 
The kicked rotor was first realized in the laboratory by exposing ultracold sodium atoms to a periodic sequence of sharp pulses of near-resonant light~\cite{raizen}.
Signatures of classical and quantum chaotic behavior, including momentum diffusion, dynamical localization, and quantum resonances have all been observed in such atom optics experiments. 
Intriguing connections have also been demonstrated between momentum localization in the quantum kicked rotor and Anderson localization in disordered lattices~\cite{fishman}.
The use of short pulses for the purpose of control of quantum systems was suggested previously for a variety of systems, including excitation of electronic states in molecules~\cite{Kosloff92}, product formation in chemical reactions~\cite{KosloffRice89, Rabitz88}, and quantum computing~\cite{PalaoKosloff02}.
Due to complexity of those systems, control pulses have to be carefully shaped to achieve effective control, and determination of such shapes often requires one to use either numerical techniques, or genetic algorithms.
The response of quantum systems to fast pulses has also been studied extensively
in the context of pulsed nuclear magnetic resonance (NMR)~\cite{slichter}. In
this case, the pulse width is typically short compared with the time scales of
the relaxation processes, including spin-lattice relaxation time $T_1$ and
transverse relaxation time $T_2^\ast$ associated with line broadening. A single
pulse may be used to study free induction decay; more sophisticated multi-pulse
sequences are used in spin echo experiments and in multi-dimensional fourier
transform NMR~\cite{ernst}. Techniques similar to ones discussed here are
therefore developed to study the detailed evolution of a single spin or
multiple spins, corresponding to multi-qubit systems, including spin
precession and relaxation effects in the time intervals between pulses. Quantum
gates necessary for computation have been constructed using such pulse
sequences~\cite{jones}, and realized in experiments~\cite{vandersypen}.
In some systems, discussed in this paper, one may take advantage of single or multiple external pulses of simple shape (e.g. Gaussian) that contain many frequency components to reliably control quantum systems.
There are other limits in which analytic solutions may be obtained, including perturbative and constant external interactions, and degenerate systems.  
However, these tend to be of limited use, as we discuss below.

In this paper we develop simple analytic solutions for singly and multiply kicked
two-state quantum systems~\cite{berman}.  
Since the kicked limit is an ideal limit of pulses very sharp in time,
we do numerical calculations for pulses of finite width to illustrate the
region of validity for fast pulses.  We do this for $2s-2p$ dipole transitions
in hydrogen and illustrate the limits on the band width of the signals required
to sensibly access such a region.  
Part of our motivation for this study is to understand reaction dynamics~\cite{mcbook}
and coherent control~\cite{qcontrol} in the time domain.  
In particular we are interested in the study of observable effects
of time ordering and also in understanding the related problem
of time correlation~\cite{gm01} in few body dynamics~\cite{gm01a}, corresponding to 
a system of a few dynamically coupled qubits.  
We use both analytic and numerical solutions to 
study effects of time ordering in multiply kicked systems.  
In a simple analytic example we illustrate the difference between 
time ordering and time reversal invariance.
Atomic units are used throughout the paper.

\section{Dynamics of a two-state system}

\subsection{Basic equations}
A two-state quantum system may be described by a wave function,
$\vert \psi \rangle= a_1 \pmatrix{1 \cr 0} + a_2 \pmatrix{0 \cr 1}$,
where $\pmatrix{0 \cr 1}$ and  $\pmatrix{1 \cr 0}$ represent the two basis states,
e.g. on and off.
Here $a_1$ and $a_2$ are complex probability amplitudes restricted by the normalization condition, $|a_1|^2 +|a_2|^2 = 1$.

The basis states $\pmatrix{1 \cr 0}$ and $\pmatrix{0 \cr 1}$ are eigenstates 
of an unperturbed Hamiltonian, $\hat H_0$, given here by,
\begin{equation}
\label{H0}
\hat H_0 = -{\Delta E\over 2} \sigma_z \ ,
\end{equation}
where $\Delta E= E_2 - E_1$ is the energy difference of the eigenstates of
$\hat H_0$, and $\sigma_z$ is a Pauli spin matrix.
The average energy of `on' and `off' states of the unperturbed system may always be taken as zero since a shift in overall energy of the system corresponds to an unphysical overall phase
in the wavefunction.
Probability amplitudes evolve as $a_j(t)=a_j(0) \rme^{-\rmi E_j t}$, and the occupation probabilities of the basis states, $P_j=|a_j|^2$, remain constant in time.

The states of a qubit can be coupled via an external interaction $\hat V(t)$,
so that the occupation probabilities change in time. 
For simplicity, we assume that the interaction has the following form:
\begin{equation}
\label{V}
\hat V (t) = V(t) \sigma_x \ ,
\end{equation}
i.e. all of the time dependence in the interaction operator $\hat V(t)$ is contained 
in a single real function of $t$ 
and the interaction does not contain a term proportional to $\hat H_0$. 
Without loss of generality, in this section we only consider interactions that include terms proportional to $\sigma_x$. 
These assumptions are often justifiable on experimental grounds~\cite{zhao,aarhus,bruch}.
The Hamiltonian of the system then becomes,
\begin{equation}
\label{H}
\hat H(t) = \hat H_0 + \hat V(t) = -{\Delta E\over 2} \sigma_z + V(t) \sigma_x \ ,
\end{equation}
and the probability amplitudes evolve according to
\begin{equation}
\label{eqmo}
\rmi{\rmd \over \rmd t} \pmatrix{a_1(t) \cr a_2(t)}
 = \pmatrix{-\Delta E/2 & V(t) \cr V(t) & \Delta E/2} \pmatrix{a_1(t) \cr a_2(t)} \ .
\end{equation}

Formal solution to (\ref{eqmo}) may be written in terms of the time evolution
operator $\hat U(t)$ as
\begin{equation}
\label{amps}
\pmatrix{a_1(t) \cr a_2(t)} = \hat U(t) \pmatrix{a_1(0) \cr a_2(0)} \ .
\end{equation}
In general, solving (\ref{eqmo}) and (\ref{amps}) requires use of numerical methods.

The time evolution operator $\hat U(t)$ may be expressed here as
\begin{eqnarray}
\label{U}
\fl \hat U(t)= T \rme^{- \rmi \int^t_0 \hat H(t') \rmd t'} 
	= T \rme^{- \rmi \int^t_0 \left( -{\Delta E \over 2} \sigma_z + V(t') \sigma_x \right) \rmd t'} 
\\ \nonumber
	= T \sum_{n=0}^\infty \frac{(-\rmi)^n}{n!}  \int_0^t \hat H(t_n)
	\rmd t_n ... 
	\int_0^t \hat H(t_2) \rmd t_2 \int_0^t \hat H(t_1) \rmd t_1  \ .
\end{eqnarray}
The only non-trivial time dependence in $\hat U(t)$ arises from time dependent $\hat H(t)$ and time ordering $T$.  
The Dyson time ordering operator $T$ specifies that $\hat{H}(t_i) \hat{H}(t_j)$ is properly ordered: 
$$T \hat H(t_i) \hat H(t_j) = \hat H(t_i) \hat H(t_j) + 
\theta(t_j-t_i) \left [ \hat H(t_j),\hat H(t_i) \right ] .$$ 
Time ordering imposes a connection between the effects of $\hat H(t_i)$ and $\hat H(t_j)$ and leads to observable time ordering effects~\cite{zhao,aarhus,bruch}.
Since time ordering effects can be defined as the difference between a result with time ordering and the corresponding result in the limit of no time ordering, it is useful to specify carefully the limit without time ordering.
Removing time ordering corresponds to replacing $T \to 1$ in (\ref{U}).

\subsection{Analytical solutions}
In this paper we emphasize the utility of having analytic solutions, i.e. as compared to less transparent solutions obtained numerically.
Unfortunately there are a limited number of conditions under which analytic solutions may be obtained.  
In order of increasing complexity these include: 
\begin{enumerate}
\item Perturbative interactions~\cite{gw}.  Here the interaction $V(t)$ is sufficiently
	weak that the system largely remains in its initial state.  The solution 
	of (\ref{eqmo}) is trivial: 
	$$\fl \hat U(t) \simeq  \pmatrix{
\rme^{\rmi \frac{\Delta E}{2} t} & -\rmi\int \rme^{\rmi \Delta E (\frac{t}{2}-t')}V(t') \rmd t' \cr 
-\rmi\int \rme^{-\rmi \Delta E (\frac{t}{2}-t')}V(t') \rmd t' & \rme^{-\rmi \frac{\Delta E}{2}t}}.$$	
	The mathematical validity condition is that the action associated with the
	external field is small, i.e. $\int V(t) \rmd t << 1$.
\item Degenerate basis states~\cite{sm03}.  In this case the energy levels of the two unperturbed
	states are nearly the same.  For two-state systems the occupation probabilities
	are typically $\cos^2(\int V(t) \rmd t)$ and $\sin^2(\int V(t) \rmd t)$.
	Remarkably this form holds for both slowly and rapidly changing fields. 
	Validity requires that the action difference associated with free propagation of the two
	unperturbed states be small, i.e. $\Delta E t < < 1$.
\item Constant external fields.  Here the interaction $V(t)$ is a constant.
	The analytic solution, found from that for slowly changing fields given immediately below, 
	is mathematically similar to the physically distinct RWA solution~\cite{me}. 
\item Slowly changing (adiabatic) fields~\cite{shore, solov}.   The analytic solution of (\ref{eqmo}) is, 
	$$\fl \hat U(t) \simeq 
	\pmatrix{ \cos \Theta(t) + \rmi \frac{\Delta E}{\Omega(t)} \sin \Theta(t)  & 
	-2\rmi \frac{V(t)}{\Omega(t)} \sin \Theta(t) \cr  
	-2\rmi \frac{V(t)}{\Omega(t)} \sin \Theta(t) &   
	\cos \Theta(t) - \rmi \frac{\Delta E}{\Omega(t)} \sin \Theta(t) } \ \ , $$
	where $ \Omega(t) = \sqrt{(\Delta E)^2 + 4V(t)^2}$ and
	$\Theta(t) = \int_0^t \Omega(t') \rmd t'/2 $. 
	The validity condition $\dot V(t) << \Omega^2(t)$
	is sometimes difficult to achieve. 
\item RWA solutions. $V(t)$ oscillates with a frequency $\omega$ close to the resonant 
	frequency of the transition between the basis states, $\omega_0=\Delta E$. 
	The RWA expression for $\hat{U}$ is the same as the expression for slowly
	changing fields given immediately above, except that $\Theta(t) = \bar{\Omega} t$,
	where $\bar{\Omega}^2 = V^2 + (\Delta \omega)^2$.
	The RWA is valid~\cite{me} when the frequency of the external field, $\omega$, 
	is nearly the same as that of the transition frequency, $\omega_0$, i.e. 
	$\Delta \omega =\omega-\omega_0 << \omega_0$.
\item A sudden pulse~\cite{ostr} or series of sudden pulses (single or multiple kicks). 
	The basic validity condition~\cite{kick} is that the external field is sharply
	tuned in time, i.e. $\Delta E \tau << 1$, where $\tau$ is the width of the pulse.
	This condition is met when $\tau$ is relatively small.  We examine this in detail below.  
	In many experimentally accessible cases one can build an external field 
	using a combination of kicks.
\end{enumerate}

These solutions represent different scenarios, some of which can lead to a significant or even complete transfer of the population between basis states of the qubit.
The scenarios that allow complete transfer of the population are especially interesting for possible applications in the field of quantum information.
Any measurement of a superposition state of the qubit leads to the collapse of the wavefunction and results in finding a qubit in one of its basis states.
Therefore, the only states of a qubit for which one can predict the outcome of the measurement 
are the basis states themselves.
Consequently, one needs a reliable way to drive the system into one of these states.
Also, to form an arbitrary superposition state, one should be able to transfer any fraction of 
the population of the system into any of the states. 
This cannot be achieved with the perturbative scenario, for which the transfer of population 
is always incomplete, and some superposition states can never be formed.

In the stationary or adiabatically changing field scenario, completeness of the transfer is limited by the ratio $\Delta E /V$.  Transfer is incomplete unless the energy levels are 
degenerate.  In RWA, the probability amplitudes oscillate~\cite{me} with the Rabi 
frequency $\bar{\Omega}$, and completeness of the transfer can be adjusted by changing the detuning 
parameter, $\Delta \omega$.  Again, only in the limit of exact resonance, 
$\Delta \omega \rightarrow 0$, is the transfer complete.
Another technique, which is based on RWA and, when applicable, enables one to achieve almost complete transfer of the population, is STIRAP (stimulated Raman adiabatic passage~\cite{BergmannShore}). In this approach, a counter intuitive sequence of a pump pulse and a Stokes pulse is used to transfer the population via an intermediate state without losing any population due to the spontaneous decay of that state. This technique has proven to be very effective in a number of systems. The limiting factors there, however, are: i) it cannot be applied to degenerate or nearly degenerate systems, and ii) the pulses have to be applied adiabatically, which prevents one from using fast and ultrafast pulses (typical duration of pulses used in STIRAP is of the order of a nanosecond, whereas in the kick approach, the only restriction comes from the structure of the energy levels of the system, so picosecond and, in some cases, even femtosecond pulses can be used).
In degenerate qubits, even higher degree of controllability can be achieved~\cite{sm03}. 
And in some systems, a natural way to achieve a complete transfer of the population in a qubit 
is to apply a sudden pulse, or a kick.
The focus of this paper therefore is on suddenly changing pulses, i.e. kicks, where population
transfer can in some cases be complete.

\subsection{Pulses}
Kicks are an ideal limit of finite pulses, or sequence of pulses, each of some finite duration $\tau$. 
Each kick causes sudden changes in the populations of the two states.  
It is instructive to define phase angles for each individual pulse, namely,
\begin{eqnarray}
\label{angles}
\alpha &=& \int V(t) dt \ , \nonumber \\
\beta &=& \tau \Delta E/2 \ .
\end{eqnarray}
The angle $\alpha$ is a measure of the strength of the interaction $V(t)$ over 
the duration of a given pulse. 
The angle $\beta$ is a measure of the influence of $\hat H_0$ during the 
interaction interval $\tau$.  

Exact analytical solutions can be obtained in the limit of kicks when the pulse applied at the time $t=t_k$ becomes a $\delta$-function, $V(t) \to \alpha_k \delta (t-t_k)$, since the integration over time in (\ref{U}) becomes trivial.
For a more realistic case where the pulse has a finite width, one may, at best,
only obtain an approximate solution.  

\section{Sudden switching}
In this section we consider single and multiple kicks, where $V(t)$
may be described in terms of delta functions in time, i.e.
$V(t) = \sum_{k=1}^n \alpha_k \delta(t-t_k)$.
We work primarily in the interaction representation, since the solutions 
are relatively simple and there are generally
advantages with convergence in the interaction representation~\cite{converg}.
In the interaction representation the evolution operator has the general form
\begin{eqnarray}
\label{Ui}
  \hat U_I(t) &=& T \rme^{- \rmi\int^t_0 \hat V_I(t') \rmd t'} \nonumber \\
	&=& T \exp \left( - \rmi \int^t_0 
\rme^{-\rmi \frac{\Delta E}{2}t'\sigma_z} V(t') \sigma_x
\rme^{ \rmi \frac{\Delta E}{2}t'\sigma_z} \rmd t' \right) \ .
\end{eqnarray}
It's straightforward (e.g. using power series expansions) to show that 
$$\rme^{-\rmi \frac{\Delta E}{2}t\sigma_z} \sigma_x \rme^{\rmi \frac{\Delta E}{2}t\sigma_z}=\cos(\Delta E t)\sigma_x+\sin(\Delta E t)\sigma_y.$$
Introducing a unit vector $\vec{n}(t)=\{\cos(\Delta E t);\sin(\Delta E t);0\}$, (\ref{Ui}) can be written as
\begin{equation}
\label{Ui1}
  \hat U_I(t) = T \exp \left( - \rmi \int^t_0 V(t') \, \vec{n}(t')\!\cdot\!\vec{\sigma} \, \rmd t' \right) \ ,
\end{equation}
where $\vec{\sigma}=\{\sigma_x;\sigma_y;\sigma_z\}$.

As mentioned above there are relatively few cases in which analytic solutions are
available. 
Considered next is one set of such cases, namely singly and multiply kicked qubits.

\subsection{A single kick}
The basic building block is a two-state system subject to a single kick 
at time $t_k$, corresponding to $V(t) = \alpha_k  \delta(t-t_k)$.
The integration over time is trivial and the
time evolution operator in (\ref{Ui1}) becomes
\begin{equation} 
\label{Usk}
\fl \hat U_I^k(t)=\exp \left[ -\rmi \alpha_k \, \vec{n}(t_k)\!\cdot\!\vec{\sigma} \, \right] 
= \pmatrix{\cos\alpha_k & -\rmi\sin\alpha_k \rme^{-\rmi \Delta E t_k} \cr -\rmi\sin\alpha_k 
\rme^{\rmi \Delta E t_k}& \cos\alpha_k}
\end{equation}
for $t > t_k$.
Here we used the identity
$\rme^{\rmi\phi \vec{\sigma} \cdot \vec{u} } = 
\cos \phi \ \ \hat{I} + \rmi \sin \phi \ \ \vec{\sigma} \cdot \vec{u}$, where $\vec{u}$
is an arbitrary unit vector.  
Note that $\hat U_I^k(t)$ is independent of $t$ since the $\rme^{\rmi \Delta E t}$
factors, due to free propagation, are transferred from the evolution operator to
the wavefunction in the interaction representation.

Another way to evaluate $\hat U_I^k(t)$ is to use $\hat U_S^k(t)$ from the
Schr\"{o}dinger representation and to use the general relation,
$\hat U_I(t) = \rme^{\rmi \hat H_0 t} \hat U_S^k(t)$.  For a single kick $\hat U_S(t)$ 
has been previously evaluated~\cite{kick}, namely, 
$$\fl \hat U_S^k(t) = \pmatrix{ \rme^{\rmi\Delta E t/2} \cos\alpha_k & -\rmi \rme^{\rmi \Delta E(t/2 - t_k)} \sin\alpha_k 
\cr -\rmi \rme^{-\rmi \Delta E(t/2 - t_k)} \sin\alpha_k & \rme^{-\rmi\Delta E t/2} \cos\alpha_k } \ . $$
There is an explicit dependence on time in $\hat U_S^k(t)$.  
Even in this elementary example the expression for the time evolution matrix, $\hat{U}$,
is simpler in the interaction representation than in the Schr\"{o}dinger representation.

For a kicked qubit initially found in the state $\pmatrix{1 \cr 0}$,
the occupation probabilities are simply,
\begin{eqnarray}
\label{Pk1}
   P_1(t) &=& |a_1(t)|^2 = |U_{11}^k(t)|^2
	= \cos^2 \alpha_k  \ , \nonumber \\
   P_2(t) &=& |a_2(t)|^2 = |U_{21}^k(t)|^2
	= \sin^2 \alpha_k  \ .
\end{eqnarray}

When the pulse width is finite, the corrections to (\ref{Pk1}) are $O(\beta)$.  
These corrections result from the commutator of the free Hamiltonian $\hat H_0$ with
the interaction $\hat V$ during the time $\tau$ when the pulse is active.  
For example, in the case of a rectangular pulse of width $\tau$, the exact time evolution 
is given~\cite{kick} by
\begin{equation}
\hat U_I^{\rm rect} = 
\pmatrix{
\rme^{-\rmi\beta} \left( \cos \alpha' + \rmi \beta {\sin \alpha' \over \alpha'} \right) &
-\rmi \rme^{- \rmi \Delta E t_k} \alpha {\sin \alpha' \over \alpha'} \cr
-\rmi \rme^{  \rmi \Delta E t_k} \alpha {\sin \alpha' \over \alpha'} & 
\rme^{\rmi\beta} \left( \cos \alpha' - \rmi \beta {\sin \alpha' \over \alpha'} \right)
} \ ,
\end{equation}
where $\alpha'=\sqrt{\alpha^2+\beta^2}$.  To leading order in $\beta$, i.e. in
the width of the pulse, the error in the kicked approximation is given by
\begin{equation}
\label{rectcorr}
\delta \hat U_I (t)= \hat U_I^{\rm rect}-\hat U_I^k=
\rmi \beta \left( \frac{\sin \alpha}{\alpha} - \cos \alpha \right )
\sigma_z \ .
\end{equation}
In the Schr\"odinger picture, time ordering effects are present even for a single ideal kick,
specifically the time ordering between the interaction and the free evolution
preceding and following the kick~\cite{kick}. 
The time ordering effect vanishes in either the degenerate limit $\Delta E t \to 0$ or 
in the perturbative limit $\alpha \to 0$.  

In the interaction picture, time ordering effects disappear for a single ideal kick. 
This is easily understood by considering that in the interaction picture, time ordering is only 
between interactions at different times, $\hat V_I(t')$ and $\hat V_I(t'')$, not between the
interaction $\hat V(t')$ and the free Hamiltonian $\hat H_0(t'')$, as in the
Schr\"odinger case.  
For a single ideal kick, all the interaction occurs at one instant, and no ordering is needed.  
Of course, for a finite-width pulse, i.e. $\beta \ne 0$, time-ordering effects do begin to appear 
even in the interaction picture~\cite{kick}.  
We note that the time ordering effect in the interaction picture is independent of the measurement 
time $t$, though it does depend on the pulse width $\tau$ through the $\beta$ parameter.  

\subsection{Multiple kicks}
Consider now a series of kicks, $\alpha_1, \alpha_2, ... ,\alpha_n $, applied at $t=t_1,t_2, ... , t_n$ with $t_1<t_2<...<t_n$, i.e. a potential of the form $V(t)=\alpha_1 \delta (t-t_1)+\alpha_2 \delta (t-t_2)+ ... +\alpha_n \delta(t-t_n)$. In the interaction representation, this potential has the form:
$$\hat V_I(t)=(\alpha_1 \delta (t-t_1)+\alpha_2 \delta (t-t_2)+...+\alpha_n \delta(t-t_n)) \vec{n}(t) \! \cdot \! \vec{\sigma} \ . $$
The evolution operator (\ref{Ui1}) becomes
\begin{equation}
\label{Umk}
\hat U_I^{mk}=T \exp \left( - \rmi \sum_{j=1}^n \alpha_j \vec{n}(t_j) \! \cdot \! \vec{\sigma}  \right) \ .
\end{equation}
For a given order of pulses, the (\ref{Umk}) can be written as
\begin{equation}
\label{Umk0}
\hat U_I^{mk}= \exp [-\rmi\alpha_n\vec{n}(t_n) \! \cdot \! \vec{\sigma}] \times...\times \exp [-\rmi\alpha_1\vec{n}(t_1)
\! \cdot \! \vec{\sigma}] \ , \nonumber 
\end{equation}
which is a simple product of time evolution operators for single kicks.
Using (\ref{Usk}), one obtains:
\begin{eqnarray}
\label{Umk0f}
\hat U_I^{mk} &=&
\pmatrix{\cos\alpha_n & -\rmi \sin\alpha_n \rme^{-\rmi \Delta E t_n} \cr -\rmi \sin\alpha_n \rme^{\rmi \Delta E t_n}& \cos\alpha_n}
 \times ... \\  \nonumber 
& \times &
\pmatrix{\cos\alpha_1 & -\rmi \sin\alpha_1 \rme^{-\rmi \Delta E t_1} \cr -\rmi \sin\alpha_1 \rme^{\rmi \Delta E t_1}& \cos\alpha_1} \ .
\end{eqnarray}
This can be evaluated for an arbitrary combination of kicks, so the analytical expression for the final occupational probabilities for the basis states can be obtained.
As we shall explicitly demonstrate later, the order in which the kicks occur can make an
observable difference.  

\subsubsection{Two arbitrary kicks}
The simplest example of a series of arbitrary kicks is a sequence of two kicks, of strengths 
$\alpha_1$ and $\alpha_2$, applied at times $t_1$ and $t_2$.
Then (\ref{Umk0f}) is easily solved, namely,
\begin{eqnarray}
\label{U2}
\hat U_I^{(2)} &=& \hat U_I^{k_2} \times \hat U_I^{k_1} = \exp [-\rmi \alpha_2\vec{n}(t_2) \vec{\sigma}] \exp [-\rmi\alpha_1\vec{n}(t_1) \vec{\sigma}]  \nonumber  \\ 
 &=&
\pmatrix{ U_{11} & -U_{21}^* \cr U_{21}  & U_{11}^* } \ ,
\end{eqnarray}
where 
\begin{eqnarray}
\label{Uij}
U_{11} &=& \cos\alpha_1 \cos\alpha_2 - \sin\alpha_1 \sin\alpha_2\rme^{-\rmi \Delta E t_-} \ , \\ \nonumber
U_{21} &=& -\rmi \rme^{\rmi \frac{\Delta E}{2} t_+} (\cos\alpha_1 \sin\alpha_2 \rme^{\rmi \frac{\Delta E}{2} t_-} 
	+ \sin\alpha_1 \cos\alpha_2 \rme^{-\rmi \frac{\Delta E}{2} t_-}) \ . 
\end{eqnarray}
Here $t_- = t_2 - t_1$, and $t_+ = t_1 + t_2$.
In the limit $t_2 \to t_1$, (\ref{U2}) reduces to (\ref{Usk}) with $\alpha \to \alpha_1+\alpha_2$.  
Note that $[\hat U_I^{k_2} , \hat U_I^{k_1}] \neq 0$ so that the time ordering of the 
interactions is important.
In the interaction representation the expression for the matrix elements of 
$\hat{U}$ and the corresponding probability amplitudes are a little simpler
than the corresponding, physically equivalent, expressions in the Schr\"{o}dinger 
representation, which include an unnecessary explicit dependence on time.
This reflects the idea that the interaction representation takes advantage
of the known eigensolutions of $\hat H_0$.  

The algebra for doing a combination of two arbitrary kicks, one proportional
to $\sigma_y$ and the other proportional to $\sigma_x$, is very similar to
that for two arbitrary kicks proportional to $\sigma_x$.  For a single
$\sigma_y$ kick, similarly to (\ref{Usk}) one quickly finds 
$\hat U^{k_y}=\pmatrix{ \cos\alpha_k & -\rme^{-\rmi \Delta E t_k} \sin\alpha_k 
\cr \rme^{\rmi \Delta E t_k} \sin\alpha_k & \cos\alpha_k}$.  

Then using $\hat U^{k_{2x} k_{1y}} = \hat U^{k_{2x}} \times \hat U^{k_{1y}}$, one finds
that the matrix elements for a $\sigma_y$ kick at $t_1$ followed
by a $\sigma_x$ kick at $t_2$ are,
\begin{eqnarray}
\label{Uijy}
U_{11} &=& \cos\alpha_1 \cos\alpha_2 - \rmi \sin\alpha_1 \sin\alpha_2\rme^{-\rmi \Delta E t_-} \ , \\ \nonumber
U_{21} &=& \rme^{ \rmi \frac{\Delta E}{2} t_+} (\cos\alpha_2 \sin\alpha_1 \rme^{-\rmi \frac{\Delta E}{2} t_-} 
	-\rmi \sin\alpha_2 \cos\alpha_1 \rme^{\rmi \frac{\Delta E}{2} t_-}) \ . 
\end{eqnarray}
From that, the transition probability to the off state is,
\begin{eqnarray}
\label{P2xy}
P_2^{k_{2x} k_{1y}} &=& |U_{21}|^2  \\ \nonumber 
  &=& \cos^2 \alpha_1 \sin^2 \alpha_2 + \sin^2 \alpha_1 \cos^2 \alpha_2
  + \frac{1}{2} \sin 2 \alpha_1 \sin 2 \alpha_2 \sin \Delta E t_- \ .
\end{eqnarray}
If the $\sigma_x$ and $\sigma_y$ kicks are reversed, then $t_- \to -t_-$
so that $P_2^{k_{2x} k_{1y}} - P_2^{k_{2y} k_{1x}}= \sin 2 \alpha_1 \sin 2 \alpha_2 \sin \Delta E t_- $.
This difference oscillates between $\pm1$ when $\alpha_1 = \alpha_2 = \pi/4$.
This effect can be observed. 
In contrast there is no observable difference for two $\sigma_x$ kicks as
may be easily shown from (\ref{Uij}).

\subsubsection{Two identical kicks}
If the pulses for two $\sigma_x$ kicks are identical ($\alpha_1=\alpha_2=\alpha$), 
then (\ref{Uij}) simplifies further to
\begin{eqnarray}
\label{Uij1}
U_{11} &=& \cos^2\alpha - \rme^{-\rmi \Delta E t_-} \sin^2\alpha \ , \\ 
U_{21} &=& -\rmi \rme^{\rmi \frac{\Delta E}{2} t_+} \sin 2\alpha \cos \frac{\Delta E}{2} t_- \ . \nonumber
\end{eqnarray}
In the limit $t_2 \to t_1$ (\ref{Uij1}) reduces to (\ref{Usk}) with $\alpha$ doubled. 

Similarly, for two $\sigma_x$ kicks of equal magnitude but opposite sign ($\alpha_1=-\alpha_2=\alpha$) applied at times $t_1$ and $t_2$, 
\begin{eqnarray}
\label{Uka0}
U_{11} &=& \cos^2 \alpha + \rme^{-\rmi \Delta E t_-} \sin^2 \alpha \ , \\ 
U_{21} &=& -\sin 2\alpha \rme^{\rmi \frac{\Delta E}{2} t_+} \sin \frac{\Delta E}{2} t_- \ , \nonumber
\end{eqnarray}
which in the limit $t_2 \to t_1$ reduces to the identity matrix.

\subsubsection{Three arbitrary kicks}
For three arbitrary $\sigma_x$ kicks of strengths $\alpha_1$, $\alpha_2$, and $\alpha_3$,
applied at times $t_1$,  $t_2$, and $t_3$,
\begin{eqnarray}
\label{U3}
\hat U_I^{(3)} &=&  U_I^{k_3} \times  U_I^{k_2} \times U_I^{k_1}   \nonumber  \\ 
 &=&
\pmatrix{ U_{11} & -U_{21}^* \cr U_{21}  & U_{11}^* } \ ,
\end{eqnarray}
where
\begin{eqnarray}
\label{U3ij}
\fl U_{11}=\cos \alpha_1 \cos \alpha_2  \cos \alpha_3 
	- \sin \alpha_1 \sin \alpha_2 \sin \alpha_3  \\ \nonumber 
	\times ( \rme^{\rmi\Delta E (t_2 - t_3)} \cot \alpha_1 
	+ \rme^{\rmi\Delta E (t_1 - t_3)} \cot \alpha_2 + \rme^{\rmi\Delta E (t_1 - t_2)} \cot \alpha_3 ) \ , 
	\\ \nonumber
\fl U_{21}=\rmi(\sin \alpha_1 \sin \alpha_2 \sin \alpha_3 \rme^{\rmi \Delta E (t_3 - t_2 + t_1)} 
	- \cos \alpha_1 \cos \alpha_2  \cos \alpha_3  \\ \nonumber
	\times (\rme^{\rmi \Delta E t_1} \tan \alpha_1   
	+ \rme^{\rmi \Delta E t_2} \tan \alpha_2 + \rme^{\rmi \Delta E t_3} \tan \alpha_3)) \ .
\end{eqnarray}

\section{Calculations}
In this section we present the results of numerical calculations of occupation probabilities  for transitions in a two-state system, caused by a series of narrow Gaussian pulses of width $\tau$.
We study the effects of the order in which pulses are applied on the final occupation probabilities.
First we illustrate the results obtained in the previous section using a model two-state system. Then we present realistic calculations for $2s \to 2p$ transitions in atomic hydrogen. We also discuss in detail the applicability of a two-state model to this transition.

\subsection{A model two-state system}
Here we present the results of numerical calculations for transitions in a model two-state system.
We directly integrate (\ref{eqmo}) using a standard fourth order Runge-Kutta method. 
In our calculations we use an interaction of the form 
$V_k(t) = (\alpha_k/\sqrt{\pi}\tau) \rme^{-(\frac{t-t_k}{\tau})^2}$, 
i.e. we replace an ideal kick (delta function) by a Gaussian pulse of strength $\alpha_k$ centered at $t_k$ with width $\tau$.
When $\tau$ is small enough for the sudden, kicked approximation to hold, this should give the same results as the analytic expression for a kick above, i.e. as $\tau \to 0$, $V_k(t) \to \alpha_k \delta(t - t_k)$.

\subsubsection{Two similar kicks}
\begin{figure}
\includegraphics{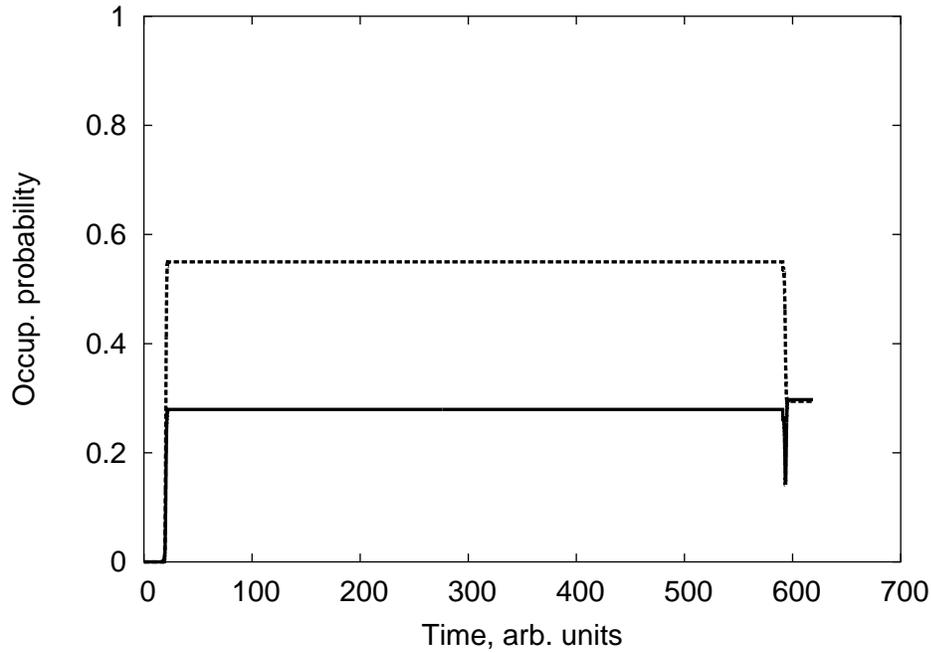}
\caption{\label{f1}
Target state probability as a function of time. Two $\sigma_x$ kicks applied at $t_1$ and $t_2$. Width of the kicks: $\tau=0.001 T_{\Delta}$. Action integral values: $\alpha_1=0.1 \pi$, $\alpha_2=0.15 \pi$ (chosen arbitrarily).
The solid line: probability for $\alpha_1$ followed by $\alpha_2$; 
the dashed line: probability for $\alpha_2$ followed by $\alpha_1$.
The sharp dip due to using a pulse of finite width is explained in the text.
The final probability doesn't depend on the order of kicks.}
\end{figure}

\begin{figure}
\includegraphics{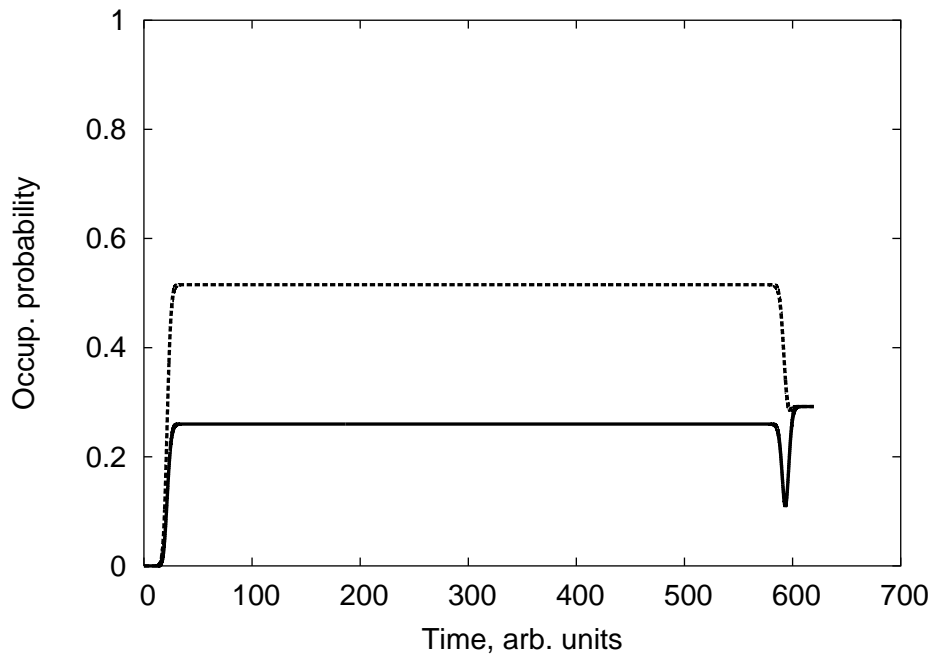}
\caption{\label{f2}
Target state probability as a function of time. Same as figure \ref{f1} but broader kicks (kick width $\tau=0.005 T_{\Delta}$).
The solid line: probability for $\alpha_1$ followed by $\alpha_2$; 
the dashed line: probability for $\alpha_2$ followed by $\alpha_1$.
The final probability still doesn't depend on the order of kicks.}
\end{figure}

First we calculate the probabilities for transitions caused by two similar kicks.
Both pulses are proportional to $\sigma_x$, but the action integral values are different: $\hat V(t)=V_1(t) \cdot \sigma_x +  V_2(t) \cdot \sigma_x$ (cf. (\ref{U2})). 
In figure \ref{f1} we show results of a calculation for the probability $P_2(t)$ 
that a system initially in the state $\pmatrix{1 \cr 0}$ makes a transition into 
the state $\pmatrix{0 \cr 1}$ when perturbed by two pulses applied at $t_1$ and $t_2$.
The ideal kick is very nearly achieved since we choose $\tau$ to be a factor of $10^{3}$ times smaller than the Rabi time, $T_{\Delta}$, for the population to oscillate between the two states.
Peaks and dips in the $P_2(t)$ graphs, that occur during application of a second pulse, reflect the following fact.
By the time second pulse is applied, the system already is in a superposition of the basic states, and it's this superposition that undergoes a precession when the pulse is on.  
The final occupation probability of the target state doesn't depend on the order in which the kicks are applied.
In figure \ref{f2}, the results of a similar calculation for broader pulses of width $\tau=0.005 T_{\Delta}$ are shown.
The shape of the $P(t)$ graph reflects the fact that the pulses have finite width.
However, the outcome of the calculation still doesn't depend on the order in which the kicks were applied.

\subsubsection{Ordering effects}
\begin{figure}
\includegraphics{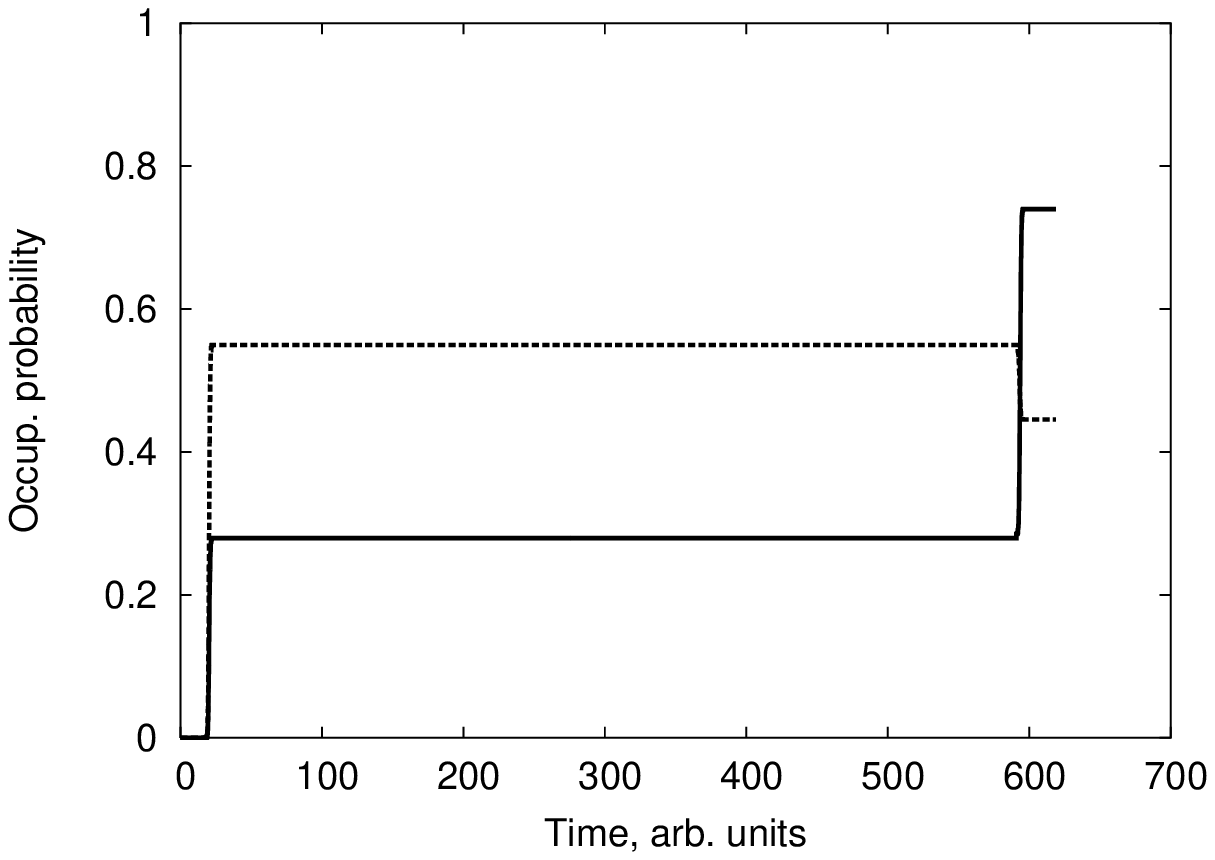}
\caption{\label{f3}
Target state probability as a function of time. Two kicks, $\alpha_1 \sigma_x$ and $\alpha_2 \sigma_y$, applied at $t_1$ and $t_2$. Kick width $\tau=0.001 T_{\Delta}$. Action integral values: $\alpha_1=0.1 \pi$, $\alpha_2=0.15 \pi$.
The solid line: probability for $\alpha_1 \sigma_x$ followed by $\alpha_2 \sigma_y$; 
the dashed line: probability for $\alpha_2 \sigma_y$ followed by $\alpha_1 \sigma_x$.
The final probability depends on the order in which the kicks are applied.}
\end{figure}

As we have shown in the previous subsection, if two pulses act on the system, then the outcome of the process is independent on the order in which they are applied as long as the pulses are similar (e.g. two $\sigma_x$ or two $\sigma_y$ pulses).
However, if the two pulses that act on the system have different structure (e.g., one is a $\sigma_x$ pulse, and the other one is a $\sigma_y$ pulse), then the results can be significantly different if different sequences of pulses are used.
In figure \ref{f3},  the occupation probability of the target state is calculated using two different sequences: an $\alpha_1 \sigma_x$ pulse followed by an $\alpha_2 \sigma_y$ pulse (solid line) and an $\alpha_2 \sigma_y$ pulse followed by an $\alpha_1 \sigma_x$ pulse (dashed line).
All the parameters are identical to the ones used in the previous part (two $\sigma_x$ pulses), except for the structure of the pulses.
The difference between two sequences is obvious. The effect of using different order of pulses, as well as the occupation probabilities for each case, are in very good agreement with the values calculated analytically (cf. (\ref{P2xy})).

\begin{figure}
\includegraphics{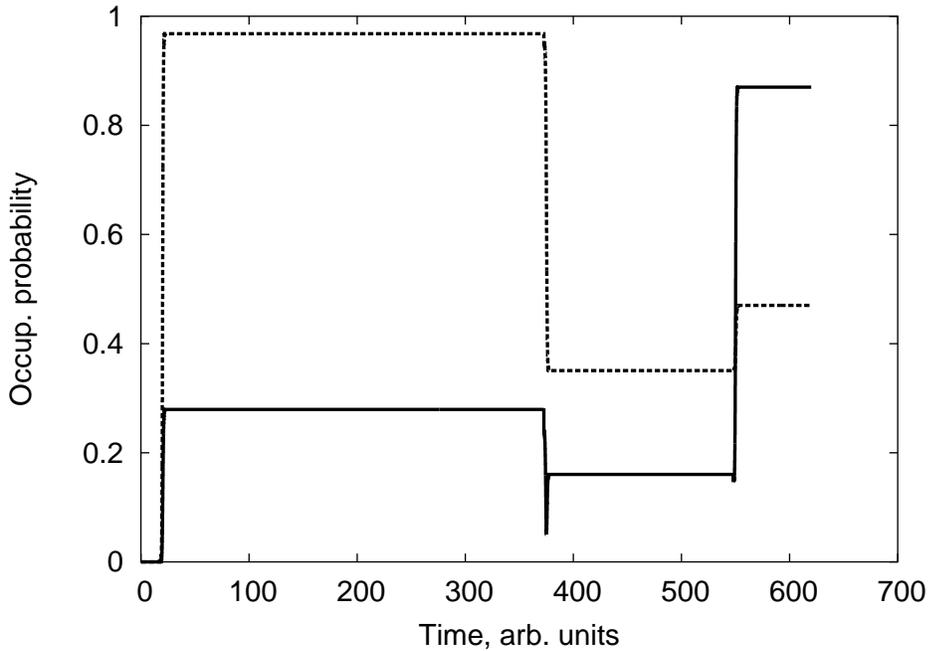}
\caption{\label{f4}
Target state probability as a function of time. Three $\sigma_x$ kicks applied at $t_1$, $t_2$, and $t_3$. Kick width $\tau=0.001 T_{\Delta}$. 
Action integral values: $\alpha_1=0.1 \pi$, $\alpha_2=0.15 \pi$, $\alpha_3=0.25 \pi$.
The solid line: probability for $\alpha_1$ followed by $\alpha_2$ followed by $\alpha_3$; 
the dashed line: probability for $\alpha_3$ followed by $\alpha_2$ followed by $\alpha_1$.
The final probability depends on the order of kicks.}
\end{figure}

Even for a sequence of pulses of the same structure, the order of pulses can be significant.
To illustrate this fact, consider a series of three $\sigma_x$ pulses.
The results of numerical calculations are shown in figure \ref{f4}.
As long as the time intervals between pulses are not the same, the outcome of the process does depend on the sequence in which the pulses are applied.

\subsection{$2s-2p$ transition in hydrogen}
In this section we present the results of numerical calculations for $2s \to 2p$ transitions in atomic hydrogen caused by a series of Gaussian pulses of width $\tau$.
Applicability of a two-state approximation to this transition is discussed in detail in the Appendix.
Specifically, we consider the fine structure splitting of the $2p$ state (target state) into $2p_{1/2}$ and $2p_{3/2}$ states. 
As we show, the phase difference accumulated during free evolution between pulses oscillates with the period $T_r=2\pi/E_{fs}$, where $E_{fs} \approx 10956$ MHz is the fine structure splitting.
Therefore, the same superposition state is formed periodically.
By choosing time intervals between kicks to be integer multiples of $T_r$, one can effectively treat the superposition of $2p_{1/2}$ and $2p_{3/2}$ states as one state ($2p$ state), that is coupled to the initial $2s$ state.
The occupation probabilities of the initial state $2s$ and the target state $2p$ are evaluated by integrating two-state equations using a standard fourth order Runge-Kutta method. 
This enables us to verify the validity of our analytic solutions for kicked qubits in the limit as $\tau \to 0$ and also to consider the effects of time ordering.

The splitting between the $2s$ and $2p_{1/2}$ states in atomic hydrogen (Lamb shift) is $\Delta E \approx 1057$ MHz. 
The corresponding time scale (the Rabi time that gives the period of oscillation between the states) is $T_{\Delta} \approx 10^{-9}$ seconds.
This gives rise to the first limitation on the duration of the pulse, $\tau$: 
it has to be significantly smaller than $T_{\Delta}$, otherwise the pulse will not be sudden and the kicked approximation will fail.  
On the other hand if $\tau$ is too small, then the interaction will have frequency components that couple the initial state to other states. 
Specifically, if $1/\tau$ is greater than $(E_{3p} - E_{2s}) \approx 10^{15}$ Hz, then the interaction will induce transitions into states with $n \ge 3$ and the system will not be well approximated by a two-state model. 
Also there is another constraint in our case. 
If the time of interaction becomes comparable to the lifetime of one of the active states (the less stable $2p$ state has a 
lifetime of $\approx 1.6 ns$), then the dissipation effects (spontaneous decay into the lower states outside two-state model) cannot be neglected.

Here we use Gaussian pulses with width $\tau=1ps$, and limit the time of interaction to $\approx 600ps$.
Single and multiple pulses of such width (and even much shorter) are achievable experimentally (e.g. half-cycle electromagnetic pulses, \cite{bucksbaum}).
We explicitly include in the numerical calculations both $2p_{1/2}$ and $2p_{3/2}$ states, and keep the time separation between the pulses $t_2-t_1=T_r$.  
The loss of population from the two-state system due to spontaneous decay ($2p \to 1s$) is also included.  
The evaluation of $\alpha$ in terms of the dipole matrix element for the $2s - 2p$ transition is discussed in a previous paper~\cite{sm03}.
We present results for the occupation probability of the target state, $P_2$, which includes both $2p_{1/2}$ and $2p_{3/2}$ states, as a function of time.

\subsubsection{Two similar kicks}
\begin{figure}
\includegraphics{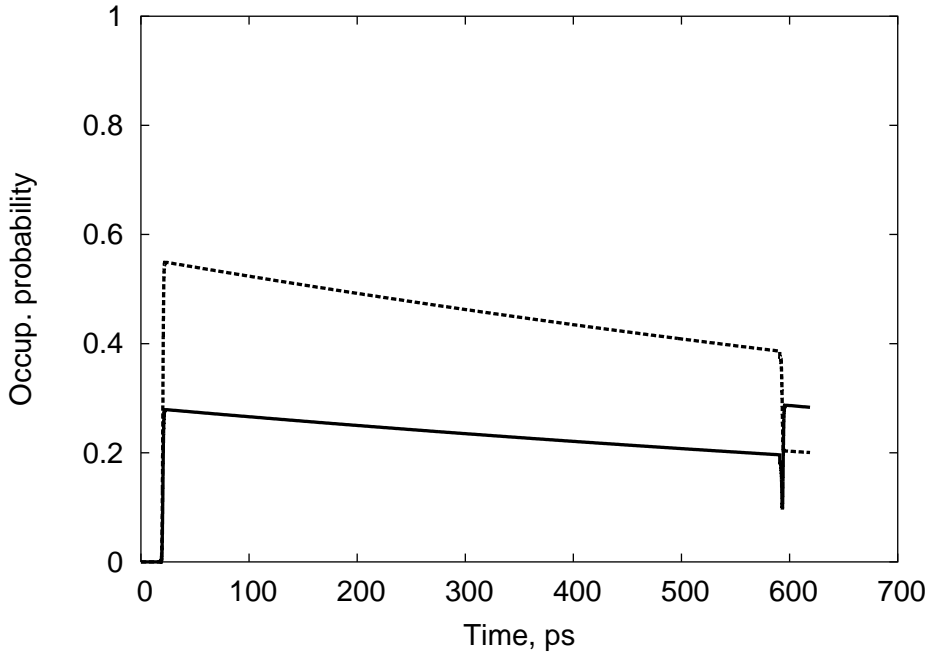}
\caption{\label{f5}
Target state probability as a function of time for $2s \to 2p$ transition in atomic hydrogen. Two $\sigma_x$ kicks applied at $t_1=20 ps$ and $t_2=593.5 ps$. Width of the kicks: $\tau=1ps$. Action integral values: $\alpha_1=0.1 \pi$, $\alpha_2=0.15 \pi$.
The solid line: probability for $\alpha_1$ followed by $\alpha_2$; 
the dashed line: probability for $\alpha_2$ followed by $\alpha_1$.
The only difference in the final probability for different orders of kicks is due to dissipation.
As in figure \ref{f1}, the sharp dips are real and due to the finite width of the pulse.}
\end{figure}

In figure \ref{f5} we show the results of a calculation for the probability $P_2(t)$ 
that a hydrogen atom  initially in the $2s$ state makes a transition into 
the $2p$ state when perturbed by two $\sigma_x$ Gaussian pulses applied at $t_1$ and $t_2$.
We have obtained our results by numerically integrating the coupled equations,
\begin{equation}
\label{2s2p-1} 
\fl \rmi \frac{\rmd}{\rmd t} \pmatrix{a_1 \cr a_2 \cr a_3}=\pmatrix{
\Delta E & -V(t) & -\sqrt{2}V(t) \cr
- V(t) & - \rmi \frac{\Gamma}{2} & 0 \cr
- \sqrt{2}V(t) & 0 & (E_{fs}- \rmi \frac{\Gamma}{2}) } \cdot \pmatrix{a_1 \cr a_2 \cr a_3} \ ,
\end{equation}
where $\Gamma \approx 626$ MHz is the decay rate for the $2p$ state, 
and we set $E_{2p_{1/2}}=0$, so $E_{2p_{1/2}}=\Delta E$ and $E_{2p_{3/2}}=E_{fs}$.
For the calculation, we used the following parameters:
kicks applied at $t_1=20 ps$ and $t_2=593.5 ps$ (separation between the pulses is equal to the revival time $T_r$ for the superposition of $2p_{1/2}$ and $2p_{3/2}$ states); action integral values: $\alpha_1=0.1 \pi$, $\alpha_2=0.15 \pi$.
The final occupation probability of the target state doesn't depend on the order in which the kicks are applied.
The small difference in the final probabilities is entirely due to dissipation effects since the decay rates of $2s$ and $2p$ states in hydrogen differ by nine orders of magnitude. 
Removing dissipation yields results indistinguishable from figure \ref{f3}.

\begin{figure}
\includegraphics{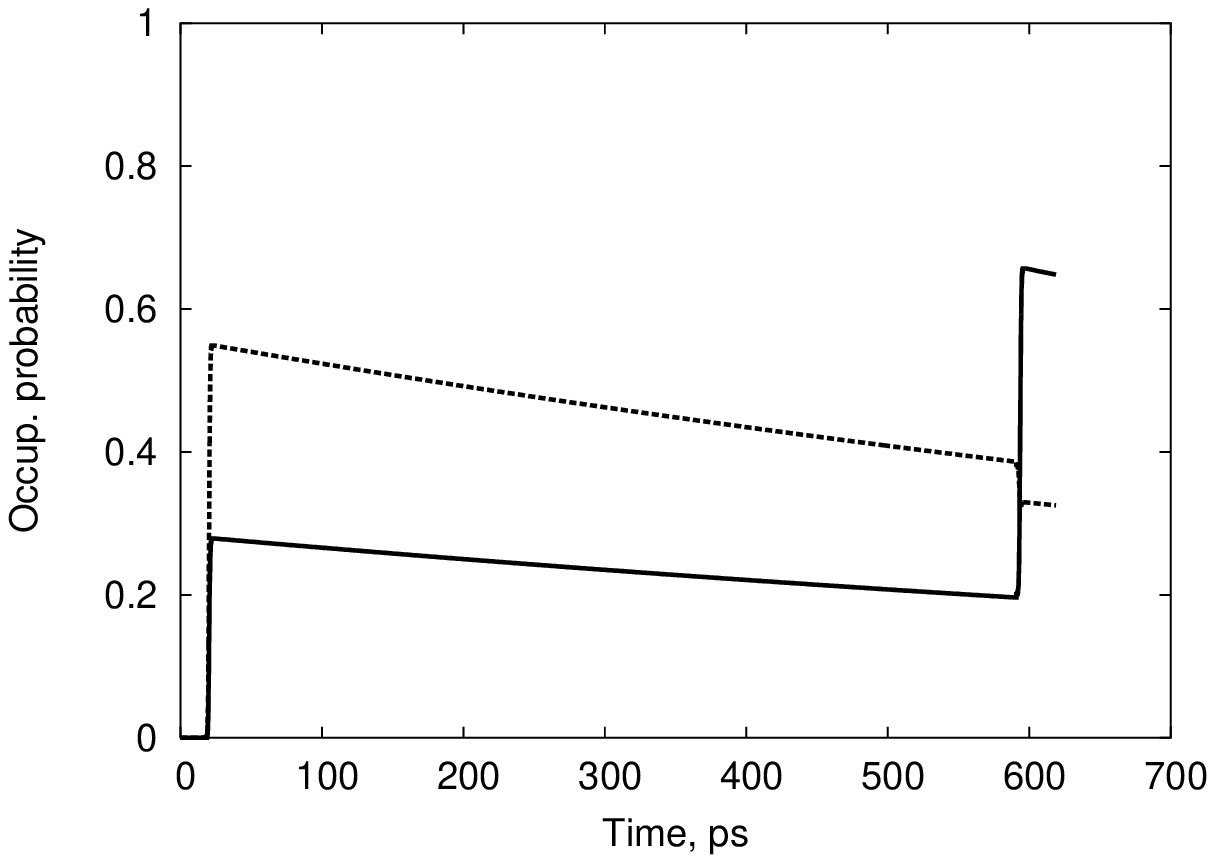}
\caption{\label{f6}
Target state probability as a function of time. Two kicks, $\alpha_1 \sigma_x$ and $\alpha_2 \sigma_y$, applied at $t_1=20 ps$ and $t_2=593.5 ps$. Kick width $\tau=1ps$. Action integral values: $\alpha_1=0.1 \pi$, $\alpha_2=0.15 \pi$.
The solid line: probability for $\alpha_1 \sigma_x$ followed by $\alpha_2 \sigma_y$; 
the dashed line: probability for $\alpha_2 \sigma_y$ followed by $\alpha_1 \sigma_x$.
The final probability depends on the order in which the kicks are applied.}
\end{figure}

In figure \ref{f6},  the occupation probability of the target state is calculated using two different sequences: an $\alpha_1 \sigma_x$ pulse followed by an $\alpha_2 \sigma_y$ pulse (solid line) and an $\alpha_2 \sigma_y$ pulse followed by an $\alpha_1 \sigma_x$ pulse (dashed line).
All the parameters are identical to the ones used in the previous part (two $\sigma_x$ pulses), except for the structure of the pulses.
Now the difference between the effects of two sequences is obvious.

\subsection{Effect of time ordering in a doubly kicked system}
In this subsection we use our analytic expressions
to examine the effect of the Dyson time ordering operator, $T$, in a kicked two-state system.
The effect of time ordering has been considered previously in the context of atomic collisions with charged particles \cite{mcbook,gm01,gm01a,zhao,aarhus,bruch} and differs somewhat from the order in which external pulses are applied, as illustrated below.
As is intuitively evident, there is no time ordering in a singly kicked
qubit~\cite{kick} since there is only one kick.  The simplest kicked
two-state system that shows an effect due to time ordering is the
qubit kicked by two equal and opposite pulses separated by a time
$t_- = t_2 - t_1$.  
The evolution matrix $\hat U_I^{-k k}$ for this system of (\ref{Uka0})
may be rewritten for convenience (as may be easily verified) as,
\begin{eqnarray}
\label{U-kk1}
\fl
\pmatrix{ 
  \rme^{-\rmi\frac{\Delta E}{2}t_-}(\cos\frac{\Delta E}{2}t_- +\rmi \cos 2 \alpha \sin \frac{\Delta E}{2}t_-)    &
  \rme^{-\rmi \Delta E t_+} \sin 2 \alpha \sin \frac{\Delta E}{2} t_-       \cr
 -\rme^{ \rmi \Delta E t_+} \sin 2 \alpha \sin \frac{\Delta E}{2} t_-  &
  \rme^{ \rmi\frac{\Delta E}{2} t_-}(\cos\frac{\Delta E}{2}t_- -\rmi \cos 2 \alpha \sin \frac{\Delta E}{2}t_-) 
} \ .
\end{eqnarray}
The limit of no time ordering, i.e.  $T \to 1$ in (\ref{Ui}),
may in principle be generally obtained~\cite{kick} by replacing $\int_0^t V_I(t') \rmd t $ with 
$ \bar{V} t $, where $ \bar{V}$ is an average (constant) value of the interaction.  In the case of two kicks of the same magnitude and opposite signs, it is then straightforward to show that, 
\begin{eqnarray}
\label{U-kk20}
\fl \hat U_{I}^{(0) -k k } =  \rme^{-\rmi \bar{V} t}  =
\pmatrix{ 
    \cos( 2 \alpha \sin \frac{\Delta E}{2} t_-)    &
  \rme^{-\rmi \Delta E t_+} \sin(2 \alpha \sin \frac{\Delta E}{2} t_-)       \cr
 -\rme^{ \rmi \Delta E t_+} \sin(2 \alpha \sin \frac{\Delta E}{2} t_-)  &
    \cos( 2 \alpha \sin \frac{\Delta E}{2} t_-)  
} \ .
\end{eqnarray}
In this example we now have analytic expressions for the matrix elements
of both $\hat U_I^{-k k}$ that contains time ordering and $\hat U_I^{(0) -k k }$
that does not include time ordering.

Let us now pause to examine the difference between time ordering and time reversal
in this simple, illustrative example.
Reversal of time ordering means that, since $\alpha_k$ and $t_k$ are
both reversed, both $t_- \to -t_-$ and $\alpha \to -\alpha$.
In this case one sees from the equations above that $\hat U_I^{-k k}$
is not invariant, i.e. phase changes occur, while $\hat U_I^{(0) -k k}$
remains the same.  Thus $\hat U_I^{-k k}$ changes when
the time ordering is changed, but $\hat U_I^{(0) -k k}$ does not change.
For time reversal~\cite{gw} $t_\pm \to -t_\pm$ and, since the initial and final states
are also interchanged, $\hat{U} \to \hat{U}^{\dag}$.
In this case one sees by inspection of the above
equations that both $\hat U_I^{-k k}$  and $\hat U_I^{(0) -k k}$ 
are invariant under time reversal, as expected.
As shown below when the symmetry of the kicks is broken
the difference between $\hat U^{k} \hat U^{k'}$ and $\hat U^{k'} \hat U^{k}$
can be observed.

We have shown above both algebraically and numerically that for two kicks
proportional to $\sigma_x$, the order of the kicks does not change
the final population transfer probability $P_2$.  However, interestingly,
this does not mean that there is no effect due to time ordering in this case.
As we show next, there is an effect due to time ordering in this case,
even though reversing the order of the kicks has no effect.
The effect of time ordering on the occupation probabilities may 
be examined by considering the probability of transfer of population from the
on state to the off state with and without time ordering, namely,

\begin{eqnarray}
\label{P2}
P_2 &=& |U_{12}|^2 = |\sin 2 \alpha  \ \sin \frac{\Delta E}{2} t_-|^2 = |\epsilon \sin \phi|^2 \ ,
 \\ \nonumber
P_2^{(0)} &=& |U_{12}^{(0)}|^2 = |\sin( 2 \alpha \sin \frac{\Delta E}{2} t_-)|^2 
	= |\sin \epsilon \phi|^2 \ \  ,
\end{eqnarray}
where $\epsilon = \sin \frac{\Delta E}{2} t_-$ and $\phi = 2 \alpha$.  

\begin{figure}
\includegraphics{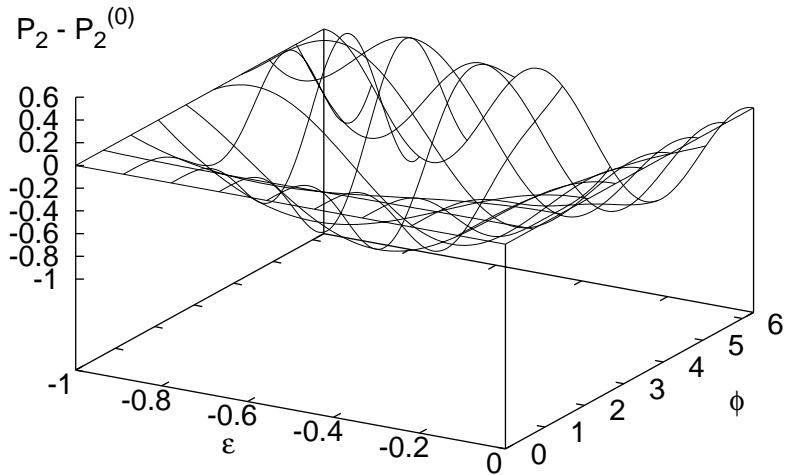}
\caption{\label{f7}
Difference in population transfer probability, $P_2 - P_2^{(0)}$ vs. 
$\epsilon = \sin \frac{\Delta E}{2} t_-$ and $\phi = 2 \alpha$.  Here $t_- = t_2 - t_1$ is
the time between the pulses and $\alpha = \int V(t') \rmd t'$ is a measure of the
interaction strength.  The two-state system is kicked by a sharp pulse of strength $\alpha$
at time $t_1$ and by an equal and opposite pulse at time $t_2$.  The difference,
$P_2 - P_2^{(0)}$, is due to time ordering in this qubit.}
\end{figure}

The effect of time ordering is shown in figure \ref{f7}, where  $P_2 - P_2^{(0)}$
is plotted as a function of $\phi = 2 \alpha$, corresponding to the strength
of the kicks, and $\epsilon = \sin \frac{\Delta E}{2} t_-$, which varies with the time separation
of the two kicks.
The effect of time ordering disappears in the example we present here in the limit that 
either the interaction strength or the time separation of the pulses goes to zero. 
For small, but finite, values of both the interaction strength and the time separation of the pulses, the effect of
time ordering is to reduce the probability of transition from the initially
occupied state to an initially unoccupied state.  
That is, in this regime time ordering reduces the maximum transfer of population
from one state to another.
As either of these two parameters gets sufficiently large, the effect of time 
ordering oscillates with increasing values of the interaction strength or
the separation time between the two pulses.
Time ordering effects are present even though $\hat U^{-kk} = \hat U^{k-k}$.

\section{Discussion}
Clearly one may extend this approach past two kicks or three.
Since arbitrary pulses can be built from a series of kicks,
in principle one may build arbitrarily complex pulses using kicks.
While adding more kicks is straightforward, the algebra becomes more difficult.
Also, the number of natural systems in which the validity conditions
apply diminishes as the pulse becomes more complex.
Hence it may be sensible to seek cases that have sufficient
symmetry so the analysis is both simple and applicable. 
It has been previously noted~\cite{kick}, for example, that a simple expression, corresponding to Floquet states,  
exists for a periodic series of narrow pulses.  

Part of the motivation for this paper grew out of an effort to
define correlation in time~\cite{gm01}, based on effects of time ordering.
While in principle we have found no fundamental problem with this effort,
we have found that it is often difficult to find expressions for the
time evolution matrix in the limit of no time ordering, namely $\hat U^0$.
Furthermore, $\hat U^0$ can depend on the representation used~\cite{kick}.
We also note that time ordering may occur in the degenerate limit,
e.g. when $\hat H = H_0 \sigma_z + V_1(t) \sigma_x + V_2(t) \sigma_y$.
In recent applications in collision dynamics using perturbation theory~\cite{gm01a}, 
time ordering is removed by use of degenerate states since the external interactions
do not contain more than one type of coupling. 
But in these calculations there is no difference between time ordering and time correlation.
The most reliable way to remove time ordering is replacement~\cite{kick} of the instantaneous
interaction $V(t)$ by its time averaged value, $\bar{V} = \frac{1}{t} \int_0^t V(t') \rmd t'$.

\vskip 3em
In summary, analytic solutions for two-state systems (e.g. qubits) strongly perturbed by a series of rapidly changing pulses, called `kicks',  have been developed and discussed.
Such analytic solutions provide useful physical insight, which together with more complete numerical methods~\cite{Lindblad, Redfield} may be used to solve more complex problems.
For a series of kicks the evolution matrix may be expressed
as a time ordered product of single kicks.
We have explicitly considered in detail single, double, and triple
kicks.
While there is no difference in the population transition probability
if two $\sigma_x$ kicks are interchanged, time ordering does have an observable effect.
The effect happens to be the same for both of these orderings.
If a $\sigma_x$ kick is interchanged with a $\sigma_y$ kick in a
doubly kicked system, the difference can be observed in most cases.
If three $\sigma_x$ kicks are used, different orderings can also
be observably different.
The effect of using pulses of finite widths has been studied numerically
for $2s - 2p$ transitions in atomic hydrogen.
If the pulse width is much smaller than the Rabi time of the active states,
then the analytic kicked solutions are valid.
Such pulses can be created experimentally using existing microwave sources.
The difference between time ordering and time reversal has been specified.
Time ordering can have observable effects.  Under time reversal the quantum
amplitudes are generally invariant.
Our results may be extended to an arbitrary number of kicks.
However, without simplifying symmetries, the solutions become more complex, and the 
applicability of this approach becomes more limited, as the
number of kicks increases.

\appendix
\section*{Appendix}
\setcounter{section}{1}
Here we discuss some details concerning the use of the two-state
approach for studying the $2s$ -- $2p$ transition in hydrogen. First, we note 
that the eight $n=2$ states (two $2s_{1/2}$ states, two $2p_{1/2}$ states,
and four $2p_{3/2}$ states) are nearly degenerate and well separated in
energy from states with $n \ne 2$.  Thus, smooth external pulses may easily be
chosen long enough to prevent field-induced population transfer out of the
$n=2$ subspace, and requiring us only to include the spontaneous decay rate
$\Gamma$ from $2p$ to $1s$.
Furthermore, rotational invariance around the axis of the external
electric field $\vec E$ leads to conservation of total angular momentum
component along that direction, allowing a given $2s_{1/2}$ state to couple
only to one $2p_{1/2}$ state and one $2p_{3/2}$ state.  Specifically, starting
with an initial $2s_{1/2}$ state with spin polarization at some angle $\chi$
relative to $\vec E$, the accessible subspace is spanned by the three basis
vectors
\begin{eqnarray}
\label{appbasis}
|2s\rangle &=& |\ell=0,\,m=0\rangle |\chi \rangle \nonumber \\
|2p\rangle &=& |\ell=1 \, ,m=0\rangle |\chi \rangle \\
|2p'\rangle &=& \cos{\chi \over 2}|\ell=1\,,m=+1\rangle |\downarrow \rangle
+\sin{\chi \over 2}|\ell=1\,,m=-1\rangle |\uparrow \rangle
 \nonumber
\end{eqnarray}
where the initial spin state is $|\chi \rangle = \cos{\chi \over 2}  |\uparrow
\rangle + \sin{\chi \over 2}  |\downarrow \rangle$, and both orbital angular
momentum and spin components are measured along the direction of $\vec E$.

Since the external pulse does not change the orbital angular momentum
component $m$, the external field couples only the two states $|2s\rangle$ 
and $|2p\rangle$ in the above basis, e.g. 
\begin{equation}
\label{appv}
\hat V(t)=V(t) \left(
\begin{array}{ccc} 0 & 1 & 0 \\ 1 & 0 & 0 \\ 0 & 0 & 0 \end{array}
\right )
\end{equation}
for a $\sigma_x$ pulse.
The free Hamiltonian $\hat H_0$ is diagonal in the basis of good total
angular momentum $j$. In the basis of (\ref{appbasis}),
\begin{equation}
\hat H_0=
\left(
\begin{array}{ccc} E_{2s_{1/2}}  & 0 & 0\\
 0 & {2 \over 3} E_{2p_{3/2}}+{1 \over 3} E_{2p_{1/2}} -\rmi{ \Gamma \over 2}
& {\sqrt{2} \over 3}
E_{fs} \\ 0& {\sqrt{2} \over 3} E_{fs} & {1 \over 3} E_{2p_{3/2}}+{2 \over
3} E_{2p_{1/2}} -\rmi {\Gamma \over 2}
\end{array}
\right ) \,,
\end{equation}
where $E_{fs}=E_{2p_{3/2}}-E_{2p_{1/2}}$ is the fine structure splitting,
and we take the decay rate $\Gamma$ to be the same for $2p_{1/2}$
and $2p_{3/2}$.
           
For narrow pulses, whose inverse width $1/\tau$ is large compared both with the
splitting $\Delta E$ between the $2s_{1/2}$ and $2p_{1/2}$ energies (Lamb shift) and also
$E_{fs}$ (fine structure),
the free propagation may be neglected during the time of the pulse.
Then the full propagator in the interaction representation
may be written as a product of kick operators
of the form $\rme^{+\rmi t \hat H_0 t_n}\rme^{-\rmi\int \rmd t \hat V_n(t)}\rme^{-\rmi\hat H_0 t_n}$,
where $t_n$ is the time of the $n^{th}$ kick.  Now $\rme^{-\rmi \int \rmd t \hat V_n(t)}$
is block-diagonal by construction (\ref{appv}), with the $2p'$ state
decoupled.  In between pulses, amplitude oscillates between the $2p$ and
$2p'$ states with period $T_r=2\pi/E_{fs}$.
However, if we now choose all inter-pulse spacings
to be integer multiples of this period,
\begin{equation}
\Delta t_n=t_{n+1}-t_n= m T_r  \,,
\end{equation}
then the free propagation between kicks also becomes diagonal:
\begin{equation}
\rme^{-\rmi \hat H_0\Delta t_n} =
\left(
\begin{array}{ccc} \rme^{-\rmi E_{2s_{1/2}}\Delta t_n}  & 0 & 0\\
 0 & \rme^{(-\rmi E_{2p_{1/2}}-\Gamma/2)\Delta t_n} & 0 
 \\ 0& 0 & \rme^{(-\rmi E_{2p_{1/2}}-\Gamma/2)\Delta t_n}
\end{array} 
\right ) \,,
\end{equation}
as may easily be checked explicitly by writing the above free propagator
in the $2p_{1/2}$, $2p_{3/2}$ basis and noting that the two basis vectors
acquire the same phase $\rme^{-\rmi E_{2p_{1/2}}\Delta t_n}=
\rme^{-\rmi E_{2p_{3/2}}\Delta t_n}$. Thus the $2p'$ state decouples entirely
and its occupation probability will always be zero
when we view the dynamics stroboscopically with period $T_r$
starting with the time $t_1$ of the first kick.  The three-state dynamics
therefore reduces to two-state dynamics in the $2s$, $2p$ subspace. 

Finally, as long the the inter-kick spacings are all integer multiples of
$T_r$, we may also use the two-state formulas presented in the
main body of the paper to evaluate the occupation probabilities at an
arbitrary time between kicks or after the last kick.  We need only remember
that the $2p$ probability that we compute at these arbitrary times
is the total probability for being in either the $2p$ and $2p'$ state,
or equivalently the total probability for being in either the
$2p_{1/2}$ or $2p_{3/2}$ state.

\ack{
DBU acknowledges support under NSF grant 0243473.
}

\begin{thebibliography}{99}
\bibitem{gm01}	Godunov A L and McGuire J H 2001 {\it \jpb} {\bf 34} L223. 
	Time correlation is defined as a deviation from the 
      	independent time approximation (ITA), where $\hat U(t) = \prod_j \hat U_j(t)$,
	where the $\hat{U}_j(t)$ include time ordering for each particle.
\end{thebibliography}

\end{document}